\documentclass[pdflatex,sn-mathphys-num]{sn-jnl}

\usepackage{graphicx}%
\usepackage{multirow}%
\usepackage{amsmath,amssymb,amsfonts}%
\usepackage{amsthm}%
\usepackage{mathrsfs}%
\usepackage[title]{appendix}%
\usepackage{xcolor}%
\usepackage{textcomp}%
\usepackage{manyfoot}%
\usepackage{booktabs}%
\usepackage{algorithm}%
\usepackage{algorithmicx}%
\usepackage{algpseudocode}%
\usepackage{listings}%
\usepackage{subcaption}
\usepackage{float}

\theoremstyle{thmstyleone}

\theoremstyle{thmstyletwo}%

\theoremstyle{thmstylethree}%

\begin{document}

\title[Quantum AI for Alzheimer's disease early screening]{Quantum AI for Alzheimer's disease early screening}

\author*[1]{\fnm{Giacomo} \sur{Cappiello}}\email{giacomo.cappiello@unifi.it}

\author*[2,3]{\fnm{Filippo} \sur{Caruso}}\email{filippo.caruso@unifi.it}

\affil[1]{\orgdiv{DiMaI “U. Dini”}, \orgname{Università degli Studi di Firenze}, \orgaddress{\street{Viale Morgagni 67/A}, \city{Firenze}, \postcode{50134}, \country{Italy}}}

\affil[2]{Dept. of Physics and Astronomy and LENS, Florence Univ., via Sansone 1, I-50019 Sesto Fiorentino, Italy}

\affil[3]{Istituto Nazionale di Ottica del Consiglio Nazionale delle Ricerche (CNR-INO), I-50019 Sesto Fiorentino, Italy}

\abstract{Quantum machine learning is a new research field combining quantum information science and machine learning. Quantum computing technologies appear to be particularly well-suited for addressing problems in the health sector efficiently. They have the potential to handle large datasets more effectively than classical models and offer greater transparency and interpretability for clinicians.

Alzheimer's disease is a neurodegenerative brain disorder that mostly affects elderly people, causing important cognitive impairments. It is the most common cause of dementia and it has an effect on memory, thought, learning abilities and movement control. This type of disease has no cure, consequently an early diagnosis is fundamental for reducing its impact. The analysis of handwriting can be effective for diagnosing, as many researches have conjectured. The DARWIN (Diagnosis AlzheimeR WIth haNdwriting) dataset contains handwriting samples from people affected by Alzheimer’s disease and a group of healthy people. Here we apply quantum AI to this use-case. In particular, we use this dataset to test classical methods for classification and compare their performances with the ones obtained via quantum machine learning me\-thods. We find that quantum methods generally perform better than classical methods.

Our results pave the way for future new quantum machine learning applications in early-screening diagnostics in the healthcare domain.}

\keywords{Quantum machine learning, Alzheimer's disease, supervised learning, data classification, parametrized quantum circuit}

\maketitle

\begin{figure}[h!]
    \centering
    \includegraphics[scale=.275]{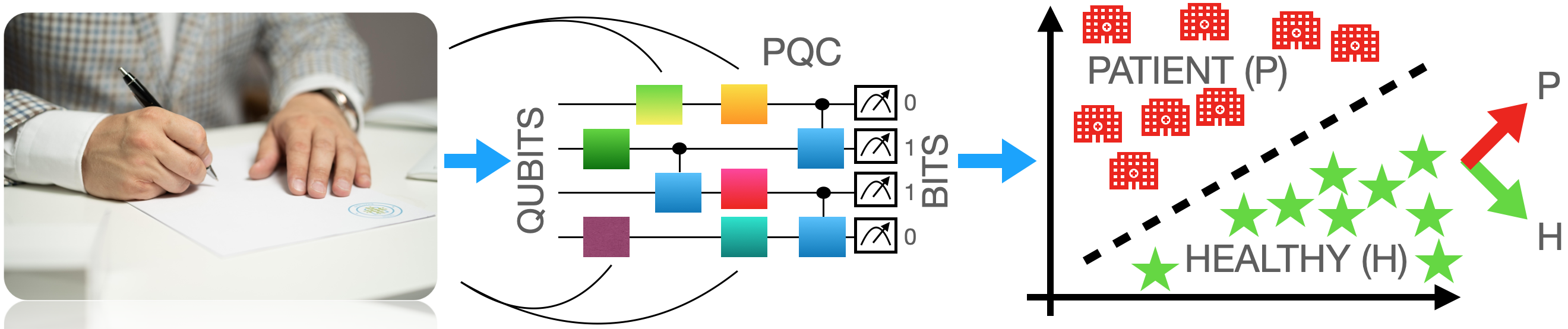}
    \caption{The data obtained from the handwriting tasks are processed and mapped into quantum states through a parametrized quantum circuit (PQC). Then a classification algorithm is applied to distinguish the samples between `patient' (P) and `healthy' (H).}
    \label{diagramma}
\end{figure}

\section{Introduction}\label{sec1}

Neurodegenerative diseases are incurable conditions caused by the progressive de\-ge\-ne\-ration of nerve cells \cite{dugger2017pathology}. Alzheimer's disease (AD) is one of the most common amongst them \cite{Honig2001-kv}. The predominant symptom in the early phase of AD is episodic memory impairment followed by progressive amnesia, a result of widespread brain damage. Since there is no cure for AD, it is critical to improve the approaches now used for diagnosis \cite{jellinger1998neuropathological}. Early diagnosis can make patients eligible for clinical trials, which are essential for the development of new treatments. Participation in these trials can provide access to cutting-edge therapies that are not yet widely available \cite{leroy2023digital} and it can help maintain cognitive function for a longer period, allowing individuals to live more independently and enjoy a better quality of life \cite{rostamzadeh2022psychotherapeutic}.

Many techniques have been used for this purpose. Neuroimaging techniques are commonly used to detect brain atrophy and other structural changes associated with AD; however their limitations include restricted availability and the inability to detect early-stage diseases \cite{ding2024speech}. Cognitive assessments, on the other hand, despite potentially being influenced by the patient’s education level, measure abilities such as memory, attention, language, and problem-solving, while motor tests examine coordination, balance, and gait. This comprehensive approach aids in fully understanding the extent of the disease’s impact \cite{mcisaac2018cognitive}. Handwriting is a task that involves both cognitive and motor functions to plan and properly execute the movements that are required with high coordination \cite{dooijes1983analysis}. The analysis of handwriting and drawing dynamics is a valid non-invasive way for e\-val\-u\-ating AD progression. For this purpose a protocol of 25 tasks was defined in Ref. \cite{CILIA2018466}. Each handwriting/drawing task is described by 18 features and the data collected amongst a group of healthy people and patients made up the DARWIN (Diagnosis AlzheimeR WIth haNdwriting) dataset. In Ref. \cite{CILIA2022104822} it is described how this dataset can be exploited by machine learning models to discriminate between AD patients and healthy people.

Machine learning (ML) is a subfield of artificial intelligence (AI) that focuses on the development of algorithms and statistical models that enable computers to learn and make predictions or decisions based on data \cite{HELM2020}. These models can identify patterns, trends, and relationships within large datasets, allowing for tasks such as classification, regression, clustering, and anomaly detection \cite{Love2002-wf}. One of the main obstacles to the usage of AI in the healthcare field, and in particular for the diagnosis of diseases, is the “black box” nature of many AI models. Medical practitioners often find it challenging to trust AI predictions because they cannot easily understand how these models arrive at their conclusions \cite{viswan2024explainable}. Quantum AI can improve the transparency and interpretability of AI models by leveraging the principles of quantum mechanics and providing a clear visual representation of models \cite{zhu2023artificial}, allowing clinicians to understand the decision-making process better. Additionally, it can handle large volumes of data more efficiently than classical AI, making it a powerful tool for data-intensive tasks \cite{ciliberto2018quantum}. In quantum machine learning (QML) inputs are encoded in quantum states and computation is done on a quantum computer \cite{SCHULD,schuld2021machine}. Although quantum processing units (QPUs) are still subjected to relatively large noise rates, it seems that applications in ML can lead to remarkable results. For example, in Ref. \cite{Jerbi2023-lo} it is shown how variational quantum models, namely models based on updating parameters in a quantum circuit through classical optimization, can exhibit a significant learning advantage in solving a regression task with input data from the fashion-MNIST dataset. A review of the contributions of QML in medical image analysis is given in Ref. \cite{WEI202342}. A pre\-vious study \cite{akpinar2023quantum} approaches via a variational quantum classifier (VQC) the problem of classification defined by the same dataset we consider in our work.

Here we focus in particular on quantum kernel methods, i.e. kernel methods in which the feature space is a space of quantum states. We utilize these algorithms to classify the samples and evaluate their performance against a classical kernel algorithm and two additional classical methods. The paper is organized as follows. In Section \ref{meto} we describe how the DARWIN dataset was created and we define the classical and quantum methods for classification. Then in Section \ref{disc} and \ref{concl} we report their performances, we discuss the results and draw our conclusions.

\section{Methods}\label{meto}

\subsection{Dataset}\label{subsec2}

The DARWIN dataset is composed of handwriting data collected according to the protocol defined in Ref. \cite{CILIA2018466}. This protocol includes 25 tasks that can be divided in the following categories: graphic tasks, like joining points and drawing geometrical figures; copy tasks, testing the ability of repeating more complex gestures to write letters, numbers and words; memory tasks, involving writing previously memorized words; and dictation tasks. The protocol was submitted to 174 participants: 89 AD patients and 85 healthy people, recruited so that the two groups would match in terms of age, education and gender. None of the participants was taking medications that influenced their cognitive abilities. The tasks were performed on paper sheets put on top of a tablet equipped with a pen that could both write in ink and sample coordinates of the tip and pressure exerted. The tablet was connected to a PC that recorded movements and displayed them in real-time. A software was used to extract 18 features for each task. Features and tasks are described in details in Appendix \ref{appen}. The dataset includes also a feature for identification of participants and another one that indicates whether the sample is associated to an AD patient (P) or an healthy person (H). This feature is used as label in the classification problem via a (classical or quantum) supervised learning model.

\subsection{Classical SVC}

\begin{figure}
    \centering
    \includegraphics[scale=0.8]{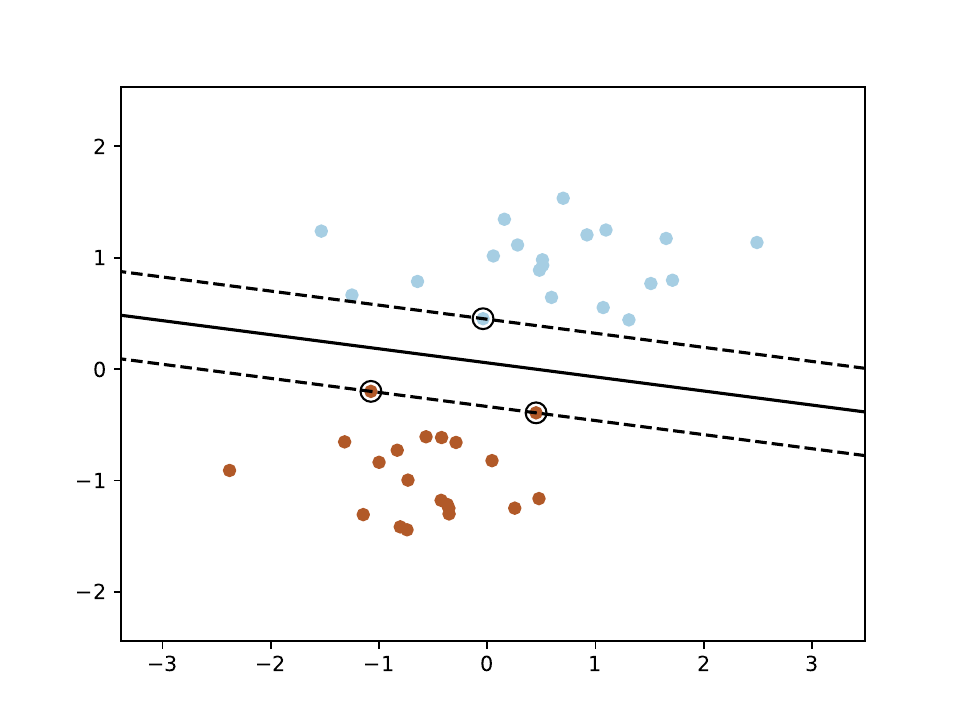}
    \caption{Graphic representation of a classification model in dimension $d=2$. The hyperplane is a straight line. Support vectors are circled.}
    \label{fig:svc}
\end{figure}

Support vector machines (SVMs) are supervised learning models for classification and regression \cite{jakkula2006tutorial}. Considering a binary classification problem in which the aim is to decide whether a set of samples belongs to one of two classes, each data point is viewed as a $d$-dimensional vector. The goal is to separate the points with a $(d-1)$-dimensional hyperplane such that the distance from the hyperplane to the nearest data points on each side, called support vectors, is maximized. A graphic representation of this is given in Fig. \ref{fig:svc}. In the original space this problem has often no solution, i.e. the sets of point are not linearly separable. A SVM uses the kernel trick, the technique of mapping inputs into high-dimensional feature spaces, to solve this problem \cite{hofmann}. The dot products of input pairs in the feature space are called kernels. Defining a map $\varphi:\mathcal{X}\rightarrow\mathcal{F}$ from the original input space to the feature space is equivalent to defining a kernel function $k:\mathcal{X}\times\mathcal{X}\rightarrow\mathbb{C}$ that maps couples of inputs into their dot products in the feature space. 

For our classification task we consider different kernel functions \cite{kavzoglu2009kernel}. The radial basis function (RBF) kernel maps a pair of inputs $(x,x')$ into $k_{rbf}(x,x')=\exp\left(-\gamma|x-x'|^2\right).$ This is one of the most common choices amongst kernel functions and it is used by default in the standard version of SVC implemented in the Scikit library in Python \cite{scikit-learn}. The linear kernel function is a simple dot product, $k_{lin}(x,x')=\langle x,x'\rangle.$ The polynomial function is $k_{pol}(x,x')=(\gamma\langle x,x'\rangle+r)^d.$ Lastly, the sigmoid kernel is given by $k_{sig}(x,x')=\tanh(\gamma\langle x,x'\rangle+r)$.

\subsection{k-Nearest Neighbors}

The k-Nearest Neighbors (kNN) model is a method used both for classification and regression \cite{kataria2013review}. With this technique, a label is assigned to a data point based on the most common label among its $k$ nearest neighbors. In particular, the probability that the unknown sample $x_0$ belongs to class $j$ is given by
\[ P(Y = j \mid X = x_0) = \frac{1}{k} \sum_{i \in N_0} I(y_i = j), \]
where $k$ is the number of nearest neighbors, $N_0$ is the set of the $k$ nearest neighbors to $x_0$ and $I(y_i = j)$ is an indicator function that equals 1 if the $i$-th neighbor $y_i$ has the label $j$, and 0 otherwise.

\subsection{Decision Tree}

A Decision Tree (DT) is a decision support tool that visually represents decisions and their possible outcomes in a hierarchical, tree-like structure \cite{kotsiantis2013decision}. This structure is composed of:
\begin{itemize}
    \item nodes, which are decision points where a specific feature is tested;
    \item branches, representing the possible outcomes of the test conducted at each node;
    \item leaf nodes where a final decision is made and a label is assigned.
\end{itemize}

The tree is constructed by recursively splitting the data based on the attribute that provides the best separation according to a chosen criterion (e.g., Gini impurity or information gain). This process continues until the stopping criteria are met, such as a maximum tree depth or a minimum number of samples per leaf node.

\subsection{Quantum SVC}

In the quantum version of SVMs, inputs are mapped into the space of quantum states \cite{schuld2021supervised}. The feature map is defined by a parametrized quantum circuit (PQC), whose parameters are given by the features of the inputs. A PQC is a sequence of parametrized unitary operators, called quantum logical gates, acting on a set of qubits, or quantum bits, which are two-state quantum systems \cite{Nielsen_Chuang_2010}. To introduce the representation of qubit states, we need to define Dirac's notation for unit vectors in $\mathbb{C}^2$. The standard basis of $\mathbb{C}^2$ is given by the vectors $\{|0\rangle,|1\rangle\},$ where $|0\rangle=\begin{pmatrix}
    1 \\ 0
\end{pmatrix}$, and 
$|1\rangle=\begin{pmatrix}
    0 \\ 1
\end{pmatrix}$. A generic unit vector in $\mathbb{C}^2$ is $|\phi\rangle=\alpha|0\rangle+\beta|1\rangle$, where $\alpha,\beta\in\mathbb{C}$ are such that $|\alpha|^2+|\beta|^2=1.$ The inner product of two unit vectors $|\phi\rangle,|\psi\rangle\in\mathbb{C}^2$ is denoted by $\langle\phi|\psi\rangle,$ while their outer product is $|\phi\rangle\langle\psi|.$ The state of a qubit is defined by a hermitian, positive semi-definite matrix $\rho\in\mathbb{C}^{2\times2}$, such that its trace is equal to 1.
The state is pure when there exists a $|\phi\rangle\in\mathbb{C}^2$ such that $\rho=|\phi\rangle\langle\phi|$, otherwise it is mixed.

To compute kernels we measure the fidelity of pairs of quantum states. Fidelity evaluates the degree of similarity of a pair of quantum states \cite{liang2019quantum}. Given a pair of pure quantum states $\rho=|\phi\rangle\langle\phi|$ and $\sigma=|\psi\rangle\langle\psi|,$ their fidelity is defined by $\mathcal{F}(\rho,\sigma)=|\langle\phi|\psi\rangle|^2$. The most used generalization of this definition for a generic pair of mixed states $\rho$ and $\sigma$ is $\mathcal{F}(\rho,\sigma)=\left(\mathrm{tr}\sqrt{\sqrt{\rho}\sigma\sqrt{\rho}}\right)^2$. The calculation of fidelities is the only computation requiring a quantum hardware. All other computations in quantum SVCs are handled by a classical hardware. 

Quantum SVC is implemented using the standard version in Scikit library with precomputed kernels. We use different PQCs, scaling the number of qubits with si\-mi\-lar ansatze. The ansatze are chosen considering the following aspects. The number of quantum gates needs to be small enough to make the time needed to perform the calculations as small as possible. Time of execution is a critical point in quantum computation, since it is often limited on real QPUs.
Given a kernel function $k$, the Gram matrix associated to it is the matrix whose entries are $G_{ij}=k(x_i,x_j),$ where $x_i$ and $x_j$ are two inputs. The eigenvalue curve of the matrix $G$ is flat when the problem is difficult to learn, whereas a desirable situation is found when the eigenvalue curve is non-flat and decays, either polynomially or exponentially fast \cite{goel2017eigenvalue}. An ansatz is expressive when it explores the space of unitaries as fully and uniformally as possible. Expressive ansatze are desirable when facing a problem with no prior knowledge of it, because of their ability to adapt to the task. However it is proved that expressive anstatze are difficult to train \cite{holmes2022connecting}, while an ansatz with low expressibility is easier to train and can achieve a better performance on the task it is specialized to solve \cite{kubler2021inductive}. One possible solution to reduce expressibility is introducing a bandwidth factor, a scaling hyperparameter that multiplies the parameters in the PQC.
The bandwidth parameter limits the reach of the feature map \cite{canatar2022bandwidth}.

The PQCs we use are composed of $R_x$ and $R_y$ rotation gates, coontrolled-z and controlled-x gates. The gates $R_x$ and $R_y$ are 1-qubit parametrized unitaries that perform a rotation of an angle given by its parameter about the x-axis and the y-axis of the Bloch sphere, respectively. The Bloch sphere is a 3-dimensional geometrical representation of the states of a qubit. The matrix representations of the rotation gates are the following: 
\[R_x(\theta)=\begin{pmatrix}
\cos\left(\frac{\theta}{2}\right) & -i\sin\left(\frac{\theta}{2}\right)\\
-i\sin\left(\frac{\theta}{2}\right) & \cos\left(\frac{\theta}{2}\right)
\end{pmatrix},\quad
R_y(\theta)=\begin{pmatrix}
\cos\left(\frac{\theta}{2}\right) & -\sin\left(\frac{\theta}{2}\right)\\
\sin\left(\frac{\theta}{2}\right) & \cos\left(\frac{\theta}{2}\right)
\end{pmatrix}.
\]

A controlled gate is a quantum operation acting on multiple qubits. One or more qubits act as controls, i.e. based on their configuration, a certain gate is applied or not to the remaining qubit, called target. The controlled-x gate is a 2-qubits controlled gate. When the control is in the state $|1\rangle$ an $X$ gate is applied to the target, where $X=\begin{pmatrix}
    0 & 1 \\ 
    1 & 0
\end{pmatrix}.$ The controlled-z is also a 2-qubits controlled gate. It is symmetric, meaning that both qubits act as control, activated by the state $|1\rangle$, and target. The matrix representations of controlled-x and controlled-z gates are
\[CX=\begin{pmatrix}
    1  & 0 & 0 & 0 \\
    0  & 1 & 0 & 0 \\
    0  & 0 & 0 & 1 \\
    0  & 0 & 1 & 0 
\end{pmatrix}, \quad 
CZ=\begin{pmatrix}
    1  & 0 & 0 & 0 \\
    0  & 1 & 0 & 0 \\
    0  & 0 & 1 & 0 \\
    0  & 0 & 0 & -1 
\end{pmatrix}.\]

A control qubit activated by the state $|1\rangle$ is represented in a circuit as $\bullet.$ 
A target qubit to which an $X$ gate is applied is denoted by $\oplus$. 
Therefore the graphic representations of controlled-x and controlled-z in the circuits are, respectively, the following.
\vspace{2mm}

\begin{figure}[H]
    \centering
    \includegraphics[scale=0.15]{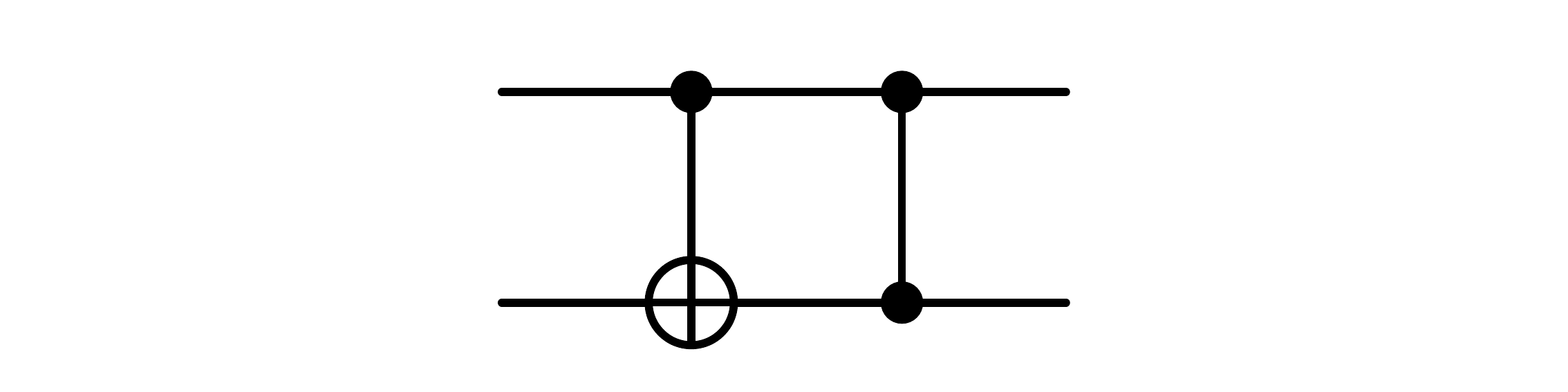}
\end{figure}

\begin{figure}[H]
    
\hspace{-1.2cm}
\includegraphics[scale=0.7]{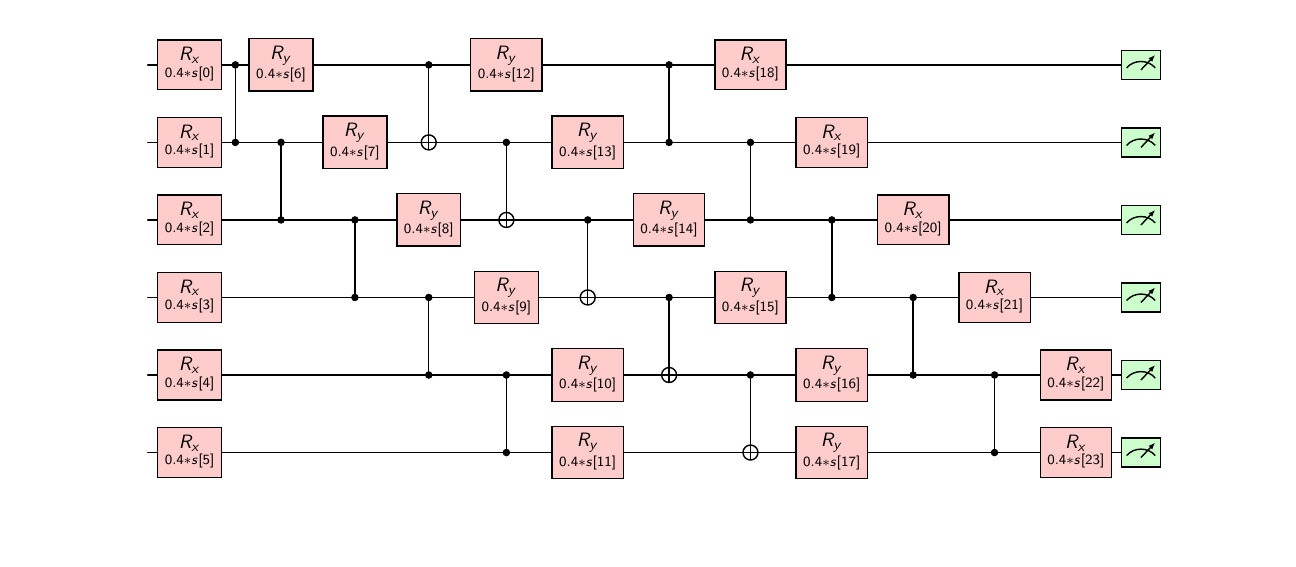}
\caption{Graphic representation of a 6-qubits PQC that maps a sample $s=(s[0],s[1],...,s[23])$ into a quantum state. The parameters in the rotation gates are given by a bandwidth factor $b=0.4$, chosen in order to maximize the accuracy, times the components of $s$.}
\label{fig:f12}
\end{figure}

In Fig. \ref{fig:f12} we show the graphic representation of a 6-qubits circuit, composed by layers of $R_x$ and $R_y$ gates alternating with layers of controlled-z and controlled-x gates. On the right-hand side are symbols representing measurements applied to the qubits, converting the quantum states in a sequence of classical bits. The number of circuit parameters is 24.

\begin{figure}[H]
    \includegraphics[scale=.8]{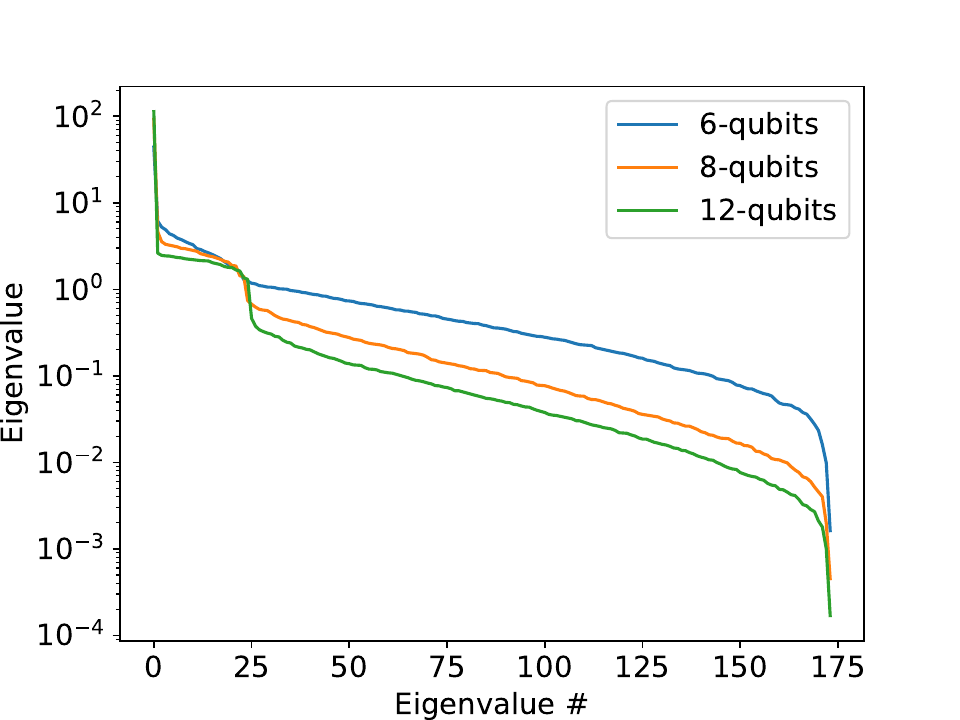}
    \caption{Plot (in logarithmic scale) of the eigenvalue curves associated with the kernel matrices defined by the 6-qubits, 8-qubits and 12-qubits circuit described in the main text.}
    \label{fig:eig}
\end{figure}

A similar structure with the same number of parameters is used to define also an 8-qubits and a 12-qubits circuit. The number of parameters must be equal to the number of features of the samples, so that each input feature corresponds to a parameter. Therefore the dataset needs to be processed through a principal component ana\-lysis (PCA) prior to applying the methods \cite{howley2005effect}. In order to reduce expressibility, the pa\-ra\-me\-ters in the rotation gates are multiplied by a tunable bandwidth hyperparameter. These circuits produce kernel matrices whose eigenvalue curve are plotted in Fig. \ref{fig:eig}. Since the curves decay, the problem can be addressed using this approach.

\subsection{Implementation}

Our algorithms are implemented in Python, using the functionalities of Scikit and Qiskit libraries \cite{Qiskit}. First we perform some preprocessing on the dataset. In particular we drop the identification feature, while keeping the feature healthy/AD's patient (H/P) as a label. We standardize the samples, by subtracting the mean and scaling to unit variance. Then we perform a PCA with 24 components. Finally we standardize the set of samples derived from the PCA. In fact, from our analysis it emerges that normalizing twice improves the performance of both classical and quantum methods. The dataset is divided for cross-validation using ShuffleSplit, performing 20 splits: 60\% of the data is used for training, 20\% for validation and 20\% as test set. For each method, a parameter grid  is defined for hyperparameter tuning: in this phase the model is trained on the training set and evaluated on the validation set. After that, the model with the best hyperparameters is trained on the union of training and validation set and is evaluated on the test set. The mean and standard deviation of the accuracies across the splits are calculated.

For classical SVC, the parameter grid includes different options for 4 hyperparameters:
\begin{itemize}
    \item `kernel', that defines the kernel function;
    \item `C', used to balance error minimization and model complexity;
    \item `gamma', which determines the influence range of training examples;
    \item `tol', which sets the precision for stopping criteria.
\end{itemize}
Two more classical methods for classifications are used for comparison, kNN and DT. For kNN the parameter grid includes 3 hyperparametrs: number of neighbors, metric and type of weights. For DT it includes 5 hyperparameters: separation criterion, splitter, maximum depth, minimum samples for split and minimum samples for leaf.

For quantum SVCs, we use fixed ansatze with 6, 8, and 12 qubits. To simplify the computations, the only hyperparameter tuned during the validation phase is the bandwidth parameter. Therefore, except for selecting the option of running the model with precomputed kernels, the other parameters of SVC() are left by default: $\mathrm{C}=1.0$, `gamma'=`scale', i.e. it scales with the number of samples, and `tol'=0.001. The quantum methods are simulated locally using Aer Simulator.

\section{Results and Discussion}\label{disc}

In Table \ref{tab:acc} we show the mean accuracy and standard deviation of classical and quantum methods. Amongst classical methods, SVC has the highest mean accuracy, but the standard deviation is relatively high, suggesting that its performance can vary significantly across different splits; kNN and DT are less effective in terms of average performance, although the latter offers more consistent results. The 6-qubits and 8-qubits SVCs perform comparably to the classical SVC, with slightly lower accuracy but a smaller standard deviation, indicating more consistent performance. The quantum SVC with 12 qubits outperforms all other models, both classical and quantum, with the highest accuracy and best consistency. This result suggests that increasing the number of qubits can enhance the performance of quantum classifiers.

\begin{table}[h]
\caption{Mean accuracy and standard deviation (expressed in percentage) of classical SVC, k-Nearest Neighbors, decision tree and quantum SVCs defined by 6-qubits, 8-qubits and 12-qubits PQCs.}
\begin{tabular*}{\textwidth}{@{\extracolsep\fill}cccccc}
\toprule
\multicolumn{3}{@{}c@{}}{Classical} & \multicolumn{3}{@{}c@{}}{Quantum} \\
\cmidrule{1-3} \cmidrule{4-6}
SVC & kNN & DT & 6q & 8q & 12q \\ \midrule
$85.28\pm 7.46$ & $69.57\pm8.20$ & $73.57\pm5.26$ & $83.57\pm 4.59$ & $83.14\pm 4.51$ & $88.29\pm 4.69$ \\ \midrule
\end{tabular*}
\label{tab:acc}
\end{table}

\subsection{Data subsampling}

Now we compare the performance of classical and quantum methods when restricted to subsets of features. For simplicity, we consider only the classical SVC and 12-qubits quantum SVC. To properly evaluate the performance of the methods on feature subsets, we use for both methods the hyperparameters that have been most frequently chosen in the previous analysis. First we use one category of features at a time, e.g. the features derived only from graphic tasks, then only copy tasks, and finally only memory and dictation tasks. The dataset is divided for cross-validation using ShuffleSplit, performing 20 splits: 80\% of the data is used for training and 20\% for testing. In Table \ref{tab:acc_sub} we show the mean accuracy and standard deviation of classical and quantum methods. 
\begin{table}[h]
\caption{Mean accuracy and standard deviation (expressed in percentage) of classical SVC and 12-qubits SVC models applied to the subset of features obtained through copy, graphic or memory and dictation tasks.}
\begin{tabular*}{\textwidth}{@{\extracolsep\fill}ccc}
\toprule
& Classical & 12q \\ \midrule
Copy & $85.57\pm8.00$ & $85.71\pm7.06$ \\ \midrule
Graphic & $78.14\pm4.97$ & $81.29\pm5.68$ \\ \midrule
Memory & $79.28\pm6.19$ & $78.57\pm4.99$ \\ \midrule
\end{tabular*}
\label{tab:acc_sub}
\end{table}

Subsequently we consider the features obtained from each of the 25 tasks in\-di\-vi\-dual\-ly. For each task we run classical SVC, store the predictions made by the 25 models and use the majority vote decision rule \cite{kittler1998combining} to select the outputs; namely for each sample we select the output predicted by most of the models. Since each task is described by 18 features, we need to define a new PQC with 18 parameters. For this purpose we choose a 9-qubits circuit. After considering all 25 tasks, we select the 5 tasks that achieve singularly the best accuracies in the previous runs. For classical SVC the tasks are 21, 17, 16, 7, 23, while for quantum SVC they are 21, 17, 24, 14, 23. As before we use the majority vote decision rule to select the predictions amongst those made by the 5 models associated with the tasks. In Table \ref{tab: by task} we show the mean accuracy and standard deviation of classical and quantum methods over 20 splittings.

\begin{table}[h]
\caption{Mean accuracy and standard deviation (expressed in percentage) of classical SVC and 9-qubits SVC models achieved combining with the majority vote decision rule the predictions made considering all tasks individually and considering only the best 5 tasks.}
\label{tab: by task}
\begin{tabular*}{\textwidth}{@{\extracolsep\fill}ccc}
\toprule
& Classical & 9q \\ \midrule
All 25 & $85.71\pm3.59$ & $86.00\pm4.30$ \\ \midrule
Best 5 & $80.28\pm4.23$ & $81.35\pm5.19$ \\ \midrule
\end{tabular*}
\end{table}

Across all data subsampling scenarios, the quantum algorithm consistently demonstrates improved or comparable performance, underscoring its robustness and potential benefits in managing diverse data types.

\subsection{Noise robustness}\label{sub: noise}

\begin{figure}
    \centering
    \includegraphics[scale=0.55]{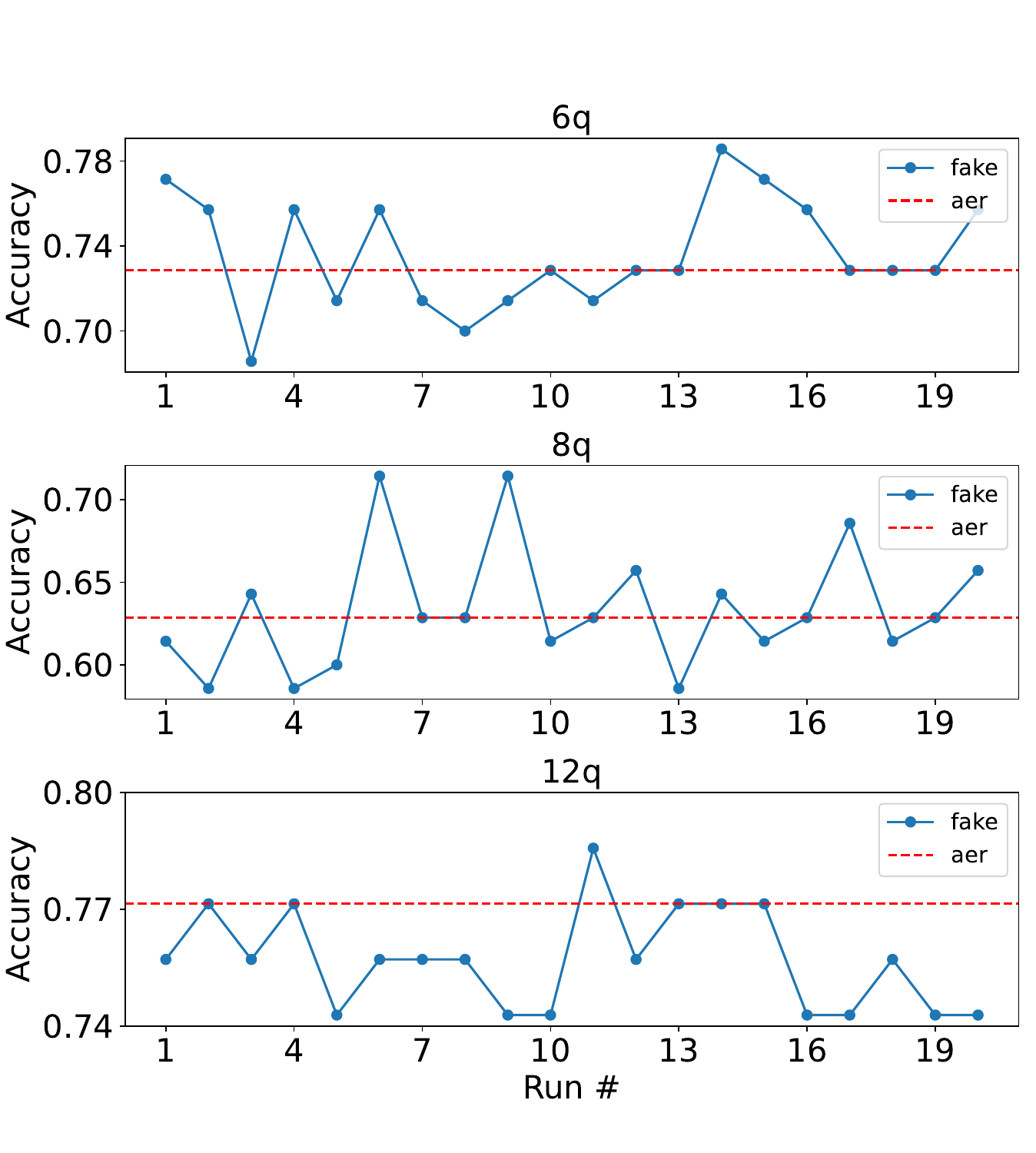}
    \caption{Accuracies of the quantum SVCs run on a noiseless simulator and on a noisy fake backend.}
    \label{fig:melb}
\end{figure}

Due to limited accessibility to real quantum hardware, computation of kernels is si\-mu\-lated without noise on a classical hardware.
Now we compare the performance of the methods executed on a noiseless Aer Simulator with those run on a fake backend. A fake backend is a simulated quantum computing environment that mimics the behavior of real IBM Quantum systems, providing real hardware qubit properties to simulate realistic conditions. We select a splitting of the dataset in training and test set, leaving 20\% of the samples for the latter, and we run the quantum SVCs on the Aer Simulator without noise, with number of shots equal to 256. This means that each circuit run is executed 256 times and the measurement results are aggregated to provide a more accurate representation of the quantum state. The default number of shots is usually set to 1024, but we choose to reduce this to speed up the computation, especially when using the fake backend. Then, with the same splitting and the same number of shots, we run the methods on Fake Melbourne backend, which simulates a 14-qubits quantum processor. The average T1 and T2 times of Fake Melbourne are 50 and 70 microseconds respectively, where T1 is a measure of how long a qubit can retain its state before it decays to ground state, and T2 indicates how long a qubit can stay in a superposition state before losing coherence, due to interactions with the environment \cite{houck2008controlling}. A noiseless backend always yields the same result, whereas the performance of a fake backend, influenced by noise, is variable. Therefore we run the methods on the fake backend 20 times, storing their predictions and using the majority rule to yield the final prediction. In Fig. \ref{fig:melb} we plot, for each of the 3 me\-thods, the accuracies achieved in the noiseless run and the 20 noisy runs. 
In Table \ref{tab: aerfake} we compare the performance of the noiseless models and the noisy models combined with the majority vote decision rule.

\begin{table}[h]
\caption{Accuracy (expressed in percentage) of quantum SVCs run on Aer Simulator without noise and on the noisy backend Fake Melbourne; for the latter the accuracy is obtained combining the predictions made on 20 executions with the majority vote decision rule.}
\label{tab: aerfake}
\begin{tabular*}{\textwidth}{@{\extracolsep\fill}cccc}
\toprule
& 6q & 8q & 12q \\ \midrule
Aer Simulator & 72.86 & 62.86 & 77.14 \\ \midrule
Fake Melbourne & 77.14 & 64.29 & 77.14 \\ \midrule
\end{tabular*}
\end{table}

The comparable or slightly better accuracy of the Fake Melbourne backend indicates that running multiple executions and using majority voting is a good strategy to mitigate variability caused by noise, achieving more consistent and accurate outcomes.

\section{Conclusion}\label{concl}

The aim of this work is comparing the performances of classical methods and quantum kernel methods for classifying samples of AD patients and healthy people subjected to handwriting tests. The classical models used are Support Vector Classification, k-Nearest Neighbors and Decision Tree; quantum SVCs are defined by ansatze of 6, 8 and 12 qubits. Among the classical methods, the SVC delivers the highest performance, albeit with significant inconsistency: average accuracy is 85.28\% and standard deviation 7.46\%. Both kNN and DT are significantly less accurate. Evaluating the quantum methods we find that 6 and 8-qubits SVCs perform similarly to the classical SVC with slightly lower accuracy but better consistency. The quantum SVC with 12 qubits emerges as the most effective model, demonstrating superior accuracy and consistency. This indicates that increasing the number of qubits can significantly boost the performance of quantum classifiers, potentially unlocking more advanced capabilities and better results.

In addition to running the methods considering all features together we have also defined methods based on data subsampling. We have considered separately the features associated with the 3 categories of tasks, namely features derived from graphic tasks, copy tasks or memory and dictation tasks. In this case we just compare the performances of the 2 best models, i.e. classical SVC and 12-qubits SVC. The quantum algorithm generally shows improved or comparable performance across different feature subsets, highlighting its robustness and potential advantages in handling diverse types of data. Then we have examined the features obtained from each of the 25 task individually, defining a model for each task, and we have used the majority vote decision rule to combine the predictions made by the 25 models; after that we have considered just the best 5 tasks, i.e. the tasks that have achieved singularly the best accuracies in the previous runs. The quantum SVC demonstrates an advantage in accuracy over the classical SVC, whether considering all tasks or just the best ones, though it tends to have higher variability in its predictions. 
 
Kernels in quantum methods have been calculated simulating quantum computation without noise. We have also investigated how noise affects the performance of quantum methods using a fake backend, which mimics the behavior of a real noisy backend. Averaging the predictions from several executions on the noisy backend, with the majority rule, leads to a slightly superior accuracy, counteracting the va\-ria\-bi\-lity introduced by noise. Several studies in classical machine learning have explored how introducing randomness and noise can enhance the performance of kernel me\-thods \cite{johnson2020kernel}\cite{Du2019}. This concept holds significant promise when applied to quantum devices, which inherently operate in noisy environments. Noise may facilitate a broader exploration of the solution space, potentially leading to better solutions that might be missed in a noise-free environment. Regarding the use of real quantum hardware, currently, only IBM Quantum systems with 127 or more qubits are accessible, and each run for our computations takes 2 to 3 seconds. Similar considerations can be made for other quantum device providers. Despite computing multiple kernels in a single run, the execution of the methods exceeds the few minutes allocated in the free plan. Additionally, the number of kernels to compute increases quadratically with the number of samples. This makes quantum kernel methods more prone to time constraints compared to other quantum classifiers that can assign a label to a sample in a single run. Exploring optimizations, such as more efficient kernel computation techniques or hybrid quantum-classical approaches, could help mitigate these time constraints and make quantum kernel methods more feasible for real-world applications.

While our experimental results are competitive with the current state of the art, there remains potential for further enhancement. In the quantum circuits we have used, only one hyperparameter, the bandwidth, is tuned in the validation phase. Enhancing the results could be achieved by incorporating one or more layers of parametric gates with tunable hyperparameters for optimization. Moreover, parameters `C' , `gamma' and `tol' could be also added in the parameter grid of quantum methods. To enhance the user-friendliness of quantum methods, particularly in the clinical diagnostic process, we aim to automate the implementation of the methods we have developed. By simplifying the process of choice of the ansatz, optimization of parameters and execution on fake or real backends, we wish to make quantum kernel methods more accessible and usable for a wide range of applications, improving the performance of classical ML models for similar problems of early screening, e.g. detection of autism in children \cite{abbas2018machine}, or identification of various types of cancer \cite{tahmooresi2018early,gould2021machine}.

\section*{Data availability}

The DARWIN dataset is available at \href{https://archive.ics.uci.edu/dataset/732/darwin}{https://archive.ics.uci.edu/dataset/732/darwin}.

\section*{Code availability}

The code used is available at request.

\bibliography{sn-article}


\begin{thebibliography}{42}
\ifx \bisbn   \undefined \def \bisbn  #1{ISBN #1}\fi
\ifx \binits  \undefined \def \binits#1{#1}\fi
\ifx \bauthor  \undefined \def \bauthor#1{#1}\fi
\ifx \batitle  \undefined \def \batitle#1{#1}\fi
\ifx \bjtitle  \undefined \def \bjtitle#1{#1}\fi
\ifx \bvolume  \undefined \def \bvolume#1{\textbf{#1}}\fi
\ifx \byear  \undefined \def \byear#1{#1}\fi
\ifx \bissue  \undefined \def \bissue#1{#1}\fi
\ifx \bfpage  \undefined \def \bfpage#1{#1}\fi
\ifx \blpage  \undefined \def \blpage #1{#1}\fi
\ifx \burl  \undefined \def \burl#1{\textsf{#1}}\fi
\ifx \doiurl  \undefined \def \doiurl#1{\url{https://doi.org/#1}}\fi
\ifx \betal  \undefined \def \betal{\textit{et al.}}\fi
\ifx \binstitute  \undefined \def \binstitute#1{#1}\fi
\ifx \binstitutionaled  \undefined \def \binstitutionaled#1{#1}\fi
\ifx \bctitle  \undefined \def \bctitle#1{#1}\fi
\ifx \beditor  \undefined \def \beditor#1{#1}\fi
\ifx \bpublisher  \undefined \def \bpublisher#1{#1}\fi
\ifx \bbtitle  \undefined \def \bbtitle#1{#1}\fi
\ifx \bedition  \undefined \def \bedition#1{#1}\fi
\ifx \bseriesno  \undefined \def \bseriesno#1{#1}\fi
\ifx \blocation  \undefined \def \blocation#1{#1}\fi
\ifx \bsertitle  \undefined \def \bsertitle#1{#1}\fi
\ifx \bsnm \undefined \def \bsnm#1{#1}\fi
\ifx \bsuffix \undefined \def \bsuffix#1{#1}\fi
\ifx \bparticle \undefined \def \bparticle#1{#1}\fi
\ifx \barticle \undefined \def \barticle#1{#1}\fi
\bibcommenthead
\ifx \bconfdate \undefined \def \bconfdate #1{#1}\fi
\ifx \botherref \undefined \def \botherref #1{#1}\fi
\ifx \url \undefined \def \url#1{\textsf{#1}}\fi
\ifx \bchapter \undefined \def \bchapter#1{#1}\fi
\ifx \bbook \undefined \def \bbook#1{#1}\fi
\ifx \bcomment \undefined \def \bcomment#1{#1}\fi
\ifx \oauthor \undefined \def \oauthor#1{#1}\fi
\ifx \citeauthoryear \undefined \def \citeauthoryear#1{#1}\fi
\ifx \endbibitem  \undefined \def \endbibitem {}\fi
\ifx \bconflocation  \undefined \def \bconflocation#1{#1}\fi
\ifx \arxivurl  \undefined \def \arxivurl#1{\textsf{#1}}\fi
\csname PreBibitemsHook\endcsname

\bibitem[\protect\citeauthoryear{Dugger and Dickson}{2017}]{dugger2017pathology}
\begin{barticle}
\bauthor{\bsnm{Dugger}, \binits{B.N.}},
\bauthor{\bsnm{Dickson}, \binits{D.W.}}:
\batitle{Pathology of neurodegenerative diseases}.
\bjtitle{Cold Spring Harbor perspectives in biology}
\bvolume{9}(\bissue{7}),
\bfpage{028035}
(\byear{2017})
\doiurl{10.1101/cshperspect.a028035}
\end{barticle}
\endbibitem

\bibitem[\protect\citeauthoryear{Honig and Mayeux}{2001}]{Honig2001-kv}
\begin{barticle}
\bauthor{\bsnm{Honig}, \binits{L.S.}},
\bauthor{\bsnm{Mayeux}, \binits{R.}}:
\batitle{Natural history of alzheimer's disease}.
\bjtitle{Aging Clinical and Experimental Research}
\bvolume{13}(\bissue{3}),
\bfpage{171}--\blpage{182}
(\byear{2001})
\doiurl{10.1007/BF03351476}
\end{barticle}
\endbibitem

\bibitem[\protect\citeauthoryear{Jellinger}{1998}]{jellinger1998neuropathological}
\begin{botherref}
\oauthor{\bsnm{Jellinger}, \binits{K.}}:
The neuropathological diagnosis of alzheimer disease.
Ageing and Dementia,
97--118
(1998)
\doiurl{10.1007/978-3-7091-6467-9_9}
\end{botherref}
\endbibitem

\bibitem[\protect\citeauthoryear{Leroy et~al.}{2023}]{leroy2023digital}
\begin{barticle}
\bauthor{\bsnm{Leroy}, \binits{V.}},
\bauthor{\bsnm{Gana}, \binits{W.}},
\bauthor{\bsnm{A{\"\i}doud}, \binits{A.}},
\bauthor{\bsnm{N’kodo}, \binits{J.-A.}},
\bauthor{\bsnm{Balageas}, \binits{A.-C.}},
\bauthor{\bsnm{Blanc}, \binits{P.}},
\bauthor{\bsnm{Bomia}, \binits{D.}},
\bauthor{\bsnm{Debacq}, \binits{C.}},
\bauthor{\bsnm{Foug{\`e}re}, \binits{B.}}:
\batitle{Digital health technologies and alzheimer’s disease clinical trials: might decentralized clinical trials increase participation by people with cognitive impairment?}
\bjtitle{Alzheimer's Research \& Therapy}
\bvolume{15}(\bissue{1}),
\bfpage{87}
(\byear{2023})
\doiurl{10.1186/s13195-023-01227-4}
\end{barticle}
\endbibitem

\bibitem[\protect\citeauthoryear{Rostamzadeh et~al.}{2022}]{rostamzadeh2022psychotherapeutic}
\begin{barticle}
\bauthor{\bsnm{Rostamzadeh}, \binits{A.}},
\bauthor{\bsnm{Kahlert}, \binits{A.}},
\bauthor{\bsnm{Kalthegener}, \binits{F.}},
\bauthor{\bsnm{Jessen}, \binits{F.}}:
\batitle{Psychotherapeutic interventions in individuals at risk for alzheimer’s dementia: a systematic review}.
\bjtitle{Alzheimer's research \& therapy}
\bvolume{14}(\bissue{1}),
\bfpage{18}
(\byear{2022})
\doiurl{10.1186/s13195-021-00956-8}
\end{barticle}
\endbibitem

\bibitem[\protect\citeauthoryear{Ding et~al.}{2024}]{ding2024speech}
\begin{barticle}
\bauthor{\bsnm{Ding}, \binits{K.}},
\bauthor{\bsnm{Chetty}, \binits{M.}},
\bauthor{\bsnm{Noori~Hoshyar}, \binits{A.}},
\bauthor{\bsnm{Bhattacharya}, \binits{T.}},
\bauthor{\bsnm{Klein}, \binits{B.}}:
\batitle{Speech based detection of alzheimer’s disease: a survey of ai techniques, datasets and challenges}.
\bjtitle{Artificial Intelligence Review}
\bvolume{57}(\bissue{12}),
\bfpage{1}--\blpage{43}
(\byear{2024})
\doiurl{10.1007/s10462-024-10961-6}
\end{barticle}
\endbibitem

\bibitem[\protect\citeauthoryear{McIsaac et~al.}{2018}]{mcisaac2018cognitive}
\begin{barticle}
\bauthor{\bsnm{McIsaac}, \binits{T.L.}},
\bauthor{\bsnm{Fritz}, \binits{N.E.}},
\bauthor{\bsnm{Quinn}, \binits{L.}},
\bauthor{\bsnm{Muratori}, \binits{L.M.}}:
\batitle{Cognitive-motor interference in neurodegenerative disease: a narrative review and implications for clinical management}.
\bjtitle{Frontiers in psychology}
\bvolume{9},
\bfpage{2061}
(\byear{2018})
\doiurl{10.3389/fpsyg.2018.02061}
\end{barticle}
\endbibitem

\bibitem[\protect\citeauthoryear{Dooijes}{1983}]{dooijes1983analysis}
\begin{barticle}
\bauthor{\bsnm{Dooijes}, \binits{E.H.}}:
\batitle{Analysis of handwriting movements}.
\bjtitle{Acta Psychologica}
\bvolume{54}(\bissue{1-3}),
\bfpage{99}--\blpage{114}
(\byear{1983})
\doiurl{10.1016/0001-6918(83)90026-4}
\end{barticle}
\endbibitem

\bibitem[\protect\citeauthoryear{Cilia et~al.}{2018}]{CILIA2018466}
\begin{barticle}
\bauthor{\bsnm{Cilia}, \binits{N.D.}},
\bauthor{\bsnm{Stefano}, \binits{C.D.}},
\bauthor{\bsnm{Fontanella}, \binits{F.}},
\bauthor{\bsnm{{Di Freca}}, \binits{A.S.}}:
\batitle{An experimental protocol to support cognitive impairment diagnosis by using handwriting analysis}.
\bjtitle{Procedia Computer Science}
\bvolume{141},
\bfpage{466}--\blpage{471}
(\byear{2018})
\doiurl{10.1016/j.procs.2018.10.141} .
\bcomment{The 9th International Conference on Emerging Ubiquitous Systems and Pervasive Networks (EUSPN-2018) / The 8th International Conference on Current and Future Trends of Information and Communication Technologies in Healthcare (ICTH-2018) / Affiliated Workshops}
\end{barticle}
\endbibitem

\bibitem[\protect\citeauthoryear{Cilia et~al.}{2022}]{CILIA2022104822}
\begin{barticle}
\bauthor{\bsnm{Cilia}, \binits{N.D.}},
\bauthor{\bsnm{{De Gregorio}}, \binits{G.}},
\bauthor{\bsnm{{De Stefano}}, \binits{C.}},
\bauthor{\bsnm{Fontanella}, \binits{F.}},
\bauthor{\bsnm{Marcelli}, \binits{A.}},
\bauthor{\bsnm{Parziale}, \binits{A.}}:
\batitle{Diagnosing alzheimer’s disease from on-line handwriting: A novel dataset and performance benchmarking}.
\bjtitle{Engineering Applications of Artificial Intelligence}
\bvolume{111},
\bfpage{104822}
(\byear{2022})
\doiurl{10.1016/j.engappai.2022.104822}
\end{barticle}
\endbibitem

\bibitem[\protect\citeauthoryear{Helm et~al.}{2020}]{HELM2020}
\begin{barticle}
\bauthor{\bsnm{Helm}, \binits{J.M.}},
\bauthor{\bsnm{Swiergosz}, \binits{A.M.}},
\bauthor{\bsnm{Haeberle}, \binits{H.S.}},
\bauthor{\bsnm{Karnuta}, \binits{J.M.}},
\bauthor{\bsnm{Schaffer}, \binits{J.L.}},
\bauthor{\bsnm{Krebs}, \binits{V.E.}},
\bauthor{\bsnm{Spitzer}, \binits{A.I.}},
\bauthor{\bsnm{Ramkumar}, \binits{P.N.}}:
\batitle{Machine learning and artificial intelligence: Definitions, applications, and future directions}.
\bjtitle{Current Reviews in Musculoskeletal Medicine}
\bvolume{13},
\bfpage{69}--\blpage{76}
(\byear{2020})
\doiurl{10.1007/s12178-020-09600-8}
\end{barticle}
\endbibitem

\bibitem[\protect\citeauthoryear{Love}{2002}]{Love2002-wf}
\begin{barticle}
\bauthor{\bsnm{Love}, \binits{B.C.}}:
\batitle{Comparing supervised and unsupervised category learning}.
\bjtitle{Psychonomic Bulletin \& Review}
\bvolume{9}(\bissue{4}),
\bfpage{829}--\blpage{835}
(\byear{2002})
\doiurl{10.3758/BF03196342}
\end{barticle}
\endbibitem

\bibitem[\protect\citeauthoryear{Viswan et~al.}{2024}]{viswan2024explainable}
\begin{barticle}
\bauthor{\bsnm{Viswan}, \binits{V.}},
\bauthor{\bsnm{Shaffi}, \binits{N.}},
\bauthor{\bsnm{Mahmud}, \binits{M.}},
\bauthor{\bsnm{Subramanian}, \binits{K.}},
\bauthor{\bsnm{Hajamohideen}, \binits{F.}}:
\batitle{Explainable artificial intelligence in alzheimer’s disease classification: A systematic review}.
\bjtitle{Cognitive Computation}
\bvolume{16}(\bissue{1}),
\bfpage{1}--\blpage{44}
(\byear{2024})
\doiurl{10.1007/s12559-023-10192-x}
\end{barticle}
\endbibitem

\bibitem[\protect\citeauthoryear{Zhu and Yu}{2023}]{zhu2023artificial}
\begin{barticle}
\bauthor{\bsnm{Zhu}, \binits{Y.}},
\bauthor{\bsnm{Yu}, \binits{K.}}:
\batitle{Artificial intelligence (ai) for quantum and quantum for ai}.
\bjtitle{Optical and Quantum Electronics}
\bvolume{55}(\bissue{8}),
\bfpage{697}
(\byear{2023})
\doiurl{10.1007/s11082-023-04914-6}
\end{barticle}
\endbibitem

\bibitem[\protect\citeauthoryear{Ciliberto et~al.}{2018}]{ciliberto2018quantum}
\begin{barticle}
\bauthor{\bsnm{Ciliberto}, \binits{C.}},
\bauthor{\bsnm{Herbster}, \binits{M.}},
\bauthor{\bsnm{Ialongo}, \binits{A.D.}},
\bauthor{\bsnm{Pontil}, \binits{M.}},
\bauthor{\bsnm{Rocchetto}, \binits{A.}},
\bauthor{\bsnm{Severini}, \binits{S.}},
\bauthor{\bsnm{Wossnig}, \binits{L.}}:
\batitle{Quantum machine learning: a classical perspective}.
\bjtitle{Proceedings of the Royal Society A: Mathematical, Physical and Engineering Sciences}
\bvolume{474}(\bissue{2209}),
\bfpage{20170551}
(\byear{2018})
\doiurl{10.1098/rspa.2017.0551}
\end{barticle}
\endbibitem

\bibitem[\protect\citeauthoryear{Schuld et~al.}{2015}]{SCHULD}
\begin{barticle}
\bauthor{\bsnm{Schuld}, \binits{M.}},
\bauthor{\bsnm{Sinayskiy}, \binits{I.}},
\bauthor{\bsnm{Petruccione}, \binits{F.}}:
\batitle{An introduction to quantum machine learning}.
\bjtitle{Contemporary Physics}
\bvolume{56}(\bissue{2}),
\bfpage{172}--\blpage{185}
(\byear{2015})
\doiurl{10.1080/00107514.2014.964942}
\end{barticle}
\endbibitem

\bibitem[\protect\citeauthoryear{Schuld and Petruccione}{2021}]{schuld2021machine}
\begin{bbook}
\bauthor{\bsnm{Schuld}, \binits{M.}},
\bauthor{\bsnm{Petruccione}, \binits{F.}}:
\bbtitle{Machine Learning with Quantum Computers},
\bedition{2}nd edn.
\bsertitle{Quantum Science and Technology},
p. \bfpage{312}.
\bpublisher{Springer},
\blocation{978-3-030-83097-7}
(\byear{2021}).
\doiurl{10.1007/978-3-030-83098-4}
\end{bbook}
\endbibitem

\bibitem[\protect\citeauthoryear{Jerbi et~al.}{2023}]{Jerbi2023-lo}
\begin{barticle}
\bauthor{\bsnm{Jerbi}, \binits{S.}},
\bauthor{\bsnm{Fiderer}, \binits{L.J.}},
\bauthor{\bsnm{Poulsen~Nautrup}, \binits{H.}},
\bauthor{\bsnm{K{\"u}bler}, \binits{J.M.}},
\bauthor{\bsnm{Briegel}, \binits{H.J.}},
\bauthor{\bsnm{Dunjko}, \binits{V.}}:
\batitle{Quantum machine learning beyond kernel methods}.
\bjtitle{Nature Communications}
\bvolume{14}(\bissue{1}),
\bfpage{517}
(\byear{2023})
\doiurl{10.1038/s41467-023-36159-y}
\end{barticle}
\endbibitem

\bibitem[\protect\citeauthoryear{Wei et~al.}{2023}]{WEI202342}
\begin{barticle}
\bauthor{\bsnm{Wei}, \binits{L.}},
\bauthor{\bsnm{Liu}, \binits{H.}},
\bauthor{\bsnm{Xu}, \binits{J.}},
\bauthor{\bsnm{Shi}, \binits{L.}},
\bauthor{\bsnm{Shan}, \binits{Z.}},
\bauthor{\bsnm{Zhao}, \binits{B.}},
\bauthor{\bsnm{Gao}, \binits{Y.}}:
\batitle{Quantum machine learning in medical image analysis: A survey}.
\bjtitle{Neurocomputing}
\bvolume{525},
\bfpage{42}--\blpage{53}
(\byear{2023})
\doiurl{10.1016/j.neucom.2023.01.049}
\end{barticle}
\endbibitem

\bibitem[\protect\citeauthoryear{Akpinar}{2023}]{akpinar2023quantum}
\begin{barticle}
\bauthor{\bsnm{Akpinar}, \binits{E.}}:
\batitle{Quantum machine learning in the cognitive domain: Alzheimer's disease study}.
\bjtitle{arXiv preprint arXiv:2401.06697}
(\byear{2023})
\doiurl{10.48550/arXiv.2401.06697}
\end{barticle}
\endbibitem

\bibitem[\protect\citeauthoryear{Jakkula}{2006}]{jakkula2006tutorial}
\begin{barticle}
\bauthor{\bsnm{Jakkula}, \binits{V.}}:
\batitle{Tutorial on support vector machine (svm)}.
\bjtitle{School of EECS, Washington State University}
\bvolume{37}(\bissue{2.5}),
\bfpage{3}
(\byear{2006})
\end{barticle}
\endbibitem

\bibitem[\protect\citeauthoryear{Hofmann et~al.}{2008}]{hofmann}
\begin{barticle}
\bauthor{\bsnm{Hofmann}, \binits{T.}},
\bauthor{\bsnm{Sch{\"o}lkopf}, \binits{B.}},
\bauthor{\bsnm{Smola}, \binits{A.J.}}:
\batitle{{Kernel methods in machine learning}}.
\bjtitle{The Annals of Statistics}
\bvolume{36}(\bissue{3}),
\bfpage{1171}--\blpage{1220}
(\byear{2008})
\doiurl{10.1214/009053607000000677}
\end{barticle}
\endbibitem

\bibitem[\protect\citeauthoryear{Kavzoglu and Colkesen}{2009}]{kavzoglu2009kernel}
\begin{barticle}
\bauthor{\bsnm{Kavzoglu}, \binits{T.}},
\bauthor{\bsnm{Colkesen}, \binits{I.}}:
\batitle{A kernel functions analysis for support vector machines for land cover classification}.
\bjtitle{International Journal of Applied Earth Observation and Geoinformation}
\bvolume{11}(\bissue{5}),
\bfpage{352}--\blpage{359}
(\byear{2009})
\doiurl{10.1016/j.jag.2009.06.002}
\end{barticle}
\endbibitem

\bibitem[\protect\citeauthoryear{Pedregosa et~al.}{2011}]{scikit-learn}
\begin{barticle}
\bauthor{\bsnm{Pedregosa}, \binits{F.}},
\bauthor{\bsnm{Varoquaux}, \binits{G.}},
\bauthor{\bsnm{Gramfort}, \binits{A.}},
\bauthor{\bsnm{Michel}, \binits{V.}},
\bauthor{\bsnm{Thirion}, \binits{B.}},
\bauthor{\bsnm{Grisel}, \binits{O.}},
\bauthor{\bsnm{Blondel}, \binits{M.}},
\bauthor{\bsnm{Prettenhofer}, \binits{P.}},
\bauthor{\bsnm{Weiss}, \binits{R.}},
\bauthor{\bsnm{Dubourg}, \binits{V.}},
\bauthor{\bsnm{Vanderplas}, \binits{J.}},
\bauthor{\bsnm{Passos}, \binits{A.}},
\bauthor{\bsnm{Cournapeau}, \binits{D.}},
\bauthor{\bsnm{Brucher}, \binits{M.}},
\bauthor{\bsnm{Perrot}, \binits{M.}},
\bauthor{\bsnm{Duchesnay}, \binits{E.}}:
\batitle{Scikit-learn: Machine learning in {P}ython}.
\bjtitle{Journal of Machine Learning Research}
\bvolume{12},
\bfpage{2825}--\blpage{2830}
(\byear{2011})
\doiurl{10.48550/arXiv.1201.0490}
\end{barticle}
\endbibitem

\bibitem[\protect\citeauthoryear{Kataria and Singh}{2013}]{kataria2013review}
\begin{barticle}
\bauthor{\bsnm{Kataria}, \binits{A.}},
\bauthor{\bsnm{Singh}, \binits{M.}}:
\batitle{A review of data classification using k-nearest neighbour algorithm}.
\bjtitle{International Journal of Emerging Technology and Advanced Engineering}
\bvolume{3}(\bissue{6}),
\bfpage{354}--\blpage{360}
(\byear{2013})
\end{barticle}
\endbibitem

\bibitem[\protect\citeauthoryear{Kotsiantis}{2013}]{kotsiantis2013decision}
\begin{barticle}
\bauthor{\bsnm{Kotsiantis}, \binits{S.B.}}:
\batitle{Decision trees: a recent overview}.
\bjtitle{Artificial Intelligence Review}
\bvolume{39},
\bfpage{261}--\blpage{283}
(\byear{2013})
\doiurl{10.1007/s10462-011-9272-4}
\end{barticle}
\endbibitem

\bibitem[\protect\citeauthoryear{Schuld}{2021}]{schuld2021supervised}
\begin{barticle}
\bauthor{\bsnm{Schuld}, \binits{M.}}:
\batitle{Supervised quantum machine learning models are kernel methods}.
\bjtitle{arXiv preprint arXiv:2101.11020}
(\byear{2021})
\doiurl{10.48550/arXiv.2101.11020}
\end{barticle}
\endbibitem

\bibitem[\protect\citeauthoryear{Nielsen and Chuang}{2010}]{Nielsen_Chuang_2010}
\begin{bbook}
\bauthor{\bsnm{Nielsen}, \binits{M.A.}},
\bauthor{\bsnm{Chuang}, \binits{I.L.}}:
\bbtitle{Quantum Computation and Quantum Information: 10th Anniversary Edition}.
\bpublisher{Cambridge University Press},
\blocation{9781107002173}
(\byear{2010}).
\doiurl{10.1017/CBO9780511976667}
\end{bbook}
\endbibitem

\bibitem[\protect\citeauthoryear{Liang et~al.}{2019}]{liang2019quantum}
\begin{barticle}
\bauthor{\bsnm{Liang}, \binits{Y.-C.}},
\bauthor{\bsnm{Yeh}, \binits{Y.-H.}},
\bauthor{\bsnm{Mendon{\c{c}}a}, \binits{P.E.}},
\bauthor{\bsnm{Teh}, \binits{R.Y.}},
\bauthor{\bsnm{Reid}, \binits{M.D.}},
\bauthor{\bsnm{Drummond}, \binits{P.D.}}:
\batitle{Quantum fidelity measures for mixed states}.
\bjtitle{Reports on Progress in Physics}
\bvolume{82}(\bissue{7}),
\bfpage{076001}
(\byear{2019})
\doiurl{10.1088/1361-6633/ab1ca4}
\end{barticle}
\endbibitem

\bibitem[\protect\citeauthoryear{Goel and Klivans}{2017}]{goel2017eigenvalue}
\begin{botherref}
\oauthor{\bsnm{Goel}, \binits{S.}},
\oauthor{\bsnm{Klivans}, \binits{A.}}:
Eigenvalue decay implies polynomial-time learnability for neural networks.
Advances in Neural Information Processing Systems
\textbf{30}
(2017)
\doiurl{10.48550/arXiv.1708.03708}
\end{botherref}
\endbibitem

\bibitem[\protect\citeauthoryear{Holmes et~al.}{2022}]{holmes2022connecting}
\begin{barticle}
\bauthor{\bsnm{Holmes}, \binits{Z.}},
\bauthor{\bsnm{Sharma}, \binits{K.}},
\bauthor{\bsnm{Cerezo}, \binits{M.}},
\bauthor{\bsnm{Coles}, \binits{P.J.}}:
\batitle{Connecting ansatz expressibility to gradient magnitudes and barren plateaus}.
\bjtitle{PRX Quantum}
\bvolume{3}(\bissue{1}),
\bfpage{010313}
(\byear{2022})
\doiurl{10.1103/PRXQuantum.3.010313}
\end{barticle}
\endbibitem

\bibitem[\protect\citeauthoryear{K{\"u}bler et~al.}{2021}]{kubler2021inductive}
\begin{barticle}
\bauthor{\bsnm{K{\"u}bler}, \binits{J.}},
\bauthor{\bsnm{Buchholz}, \binits{S.}},
\bauthor{\bsnm{Sch{\"o}lkopf}, \binits{B.}}:
\batitle{The inductive bias of quantum kernels}.
\bjtitle{Advances in Neural Information Processing Systems}
\bvolume{34},
\bfpage{12661}--\blpage{12673}
(\byear{2021})
\doiurl{10.48550/arXiv.2106.03747}
\end{barticle}
\endbibitem

\bibitem[\protect\citeauthoryear{Canatar et~al.}{2022}]{canatar2022bandwidth}
\begin{barticle}
\bauthor{\bsnm{Canatar}, \binits{A.}},
\bauthor{\bsnm{Peters}, \binits{E.}},
\bauthor{\bsnm{Pehlevan}, \binits{C.}},
\bauthor{\bsnm{Wild}, \binits{S.M.}},
\bauthor{\bsnm{Shaydulin}, \binits{R.}}:
\batitle{Bandwidth enables generalization in quantum kernel models}.
\bjtitle{arXiv preprint arXiv:2206.06686}
(\byear{2022})
\doiurl{10.48550/arXiv.2206.06686}
\end{barticle}
\endbibitem

\bibitem[\protect\citeauthoryear{Howley et~al.}{2005}]{howley2005effect}
\begin{bchapter}
\bauthor{\bsnm{Howley}, \binits{T.}},
\bauthor{\bsnm{Madden}, \binits{M.G.}},
\bauthor{\bsnm{O’Connell}, \binits{M.-L.}},
\bauthor{\bsnm{Ryder}, \binits{A.G.}}:
\bctitle{The effect of principal component analysis on machine learning accuracy with high dimensional spectral data}.
In: \bbtitle{International Conference on Innovative Techniques and Applications of Artificial Intelligence},
pp. \bfpage{209}--\blpage{222}
(\byear{2005}).
\doiurl{10.1007/1-84628-224-1_16} .
\bcomment{Springer}
\end{bchapter}
\endbibitem

\bibitem[\protect\citeauthoryear{{Qiskit contributors}}{2023}]{Qiskit}
\begin{botherref}
\oauthor{\bsnm{{Qiskit contributors}}}:
Qiskit: An Open-source Framework for Quantum Computing
(2023).
\doiurl{10.5281/zenodo.2573505}
\end{botherref}
\endbibitem

\bibitem[\protect\citeauthoryear{Kittler et~al.}{1998}]{kittler1998combining}
\begin{barticle}
\bauthor{\bsnm{Kittler}, \binits{J.}},
\bauthor{\bsnm{Hatef}, \binits{M.}},
\bauthor{\bsnm{Duin}, \binits{R.P.}},
\bauthor{\bsnm{Matas}, \binits{J.}}:
\batitle{On combining classifiers}.
\bjtitle{IEEE transactions on pattern analysis and machine intelligence}
\bvolume{20}(\bissue{3}),
\bfpage{226}--\blpage{239}
(\byear{1998})
\doiurl{10.1109/34.667881}
\end{barticle}
\endbibitem

\bibitem[\protect\citeauthoryear{Houck et~al.}{2008}]{houck2008controlling}
\begin{barticle}
\bauthor{\bsnm{Houck}, \binits{A.A.}},
\bauthor{\bsnm{Schreier}, \binits{J.}},
\bauthor{\bsnm{Johnson}, \binits{B.}},
\bauthor{\bsnm{Chow}, \binits{J.}},
\bauthor{\bsnm{Koch}, \binits{J.}},
\bauthor{\bsnm{Gambetta}, \binits{J.}},
\bauthor{\bsnm{Schuster}, \binits{D.}},
\bauthor{\bsnm{Frunzio}, \binits{L.}},
\bauthor{\bsnm{Devoret}, \binits{M.}},
\bauthor{\bsnm{Girvin}, \binits{S.}}, \betal:
\batitle{Controlling the spontaneous emission of a superconducting transmon qubit}.
\bjtitle{Physical review letters}
\bvolume{101}(\bissue{8}),
\bfpage{080502}
(\byear{2008})
\doiurl{10.1103/PhysRevLett.101.080502}
\end{barticle}
\endbibitem

\bibitem[\protect\citeauthoryear{Johnson et~al.}{2020}]{johnson2020kernel}
\begin{barticle}
\bauthor{\bsnm{Johnson}, \binits{J.E.}},
\bauthor{\bsnm{Laparra}, \binits{V.}},
\bauthor{\bsnm{P{\'e}rez-Suay}, \binits{A.}},
\bauthor{\bsnm{Mahecha}, \binits{M.D.}},
\bauthor{\bsnm{Camps-Valls}, \binits{G.}}:
\batitle{Kernel methods and their derivatives: Concept and perspectives for the earth system sciences}.
\bjtitle{Plos one}
\bvolume{15}(\bissue{10}),
\bfpage{0235885}
(\byear{2020})
\doiurl{10.1371/journal.pone.0235885}
\end{barticle}
\endbibitem

\bibitem[\protect\citeauthoryear{Du and Swamy}{2019}]{Du2019}
\begin{bbook}
\bauthor{\bsnm{Du}, \binits{K.-L.}},
\bauthor{\bsnm{Swamy}, \binits{M.N.S.}}:
\bbtitle{Kernel Methods, In: Neural Networks and Statistical Learning},
pp. \bfpage{569}--\blpage{592}.
\bpublisher{Springer},
\blocation{London}
(\byear{2019}).
\doiurl{10.1007/978-1-4471-7452-3_20}
\end{bbook}
\endbibitem

\bibitem[\protect\citeauthoryear{Abbas et~al.}{2018}]{abbas2018machine}
\begin{barticle}
\bauthor{\bsnm{Abbas}, \binits{H.}},
\bauthor{\bsnm{Garberson}, \binits{F.}},
\bauthor{\bsnm{Glover}, \binits{E.}},
\bauthor{\bsnm{Wall}, \binits{D.P.}}:
\batitle{Machine learning approach for early detection of autism by combining questionnaire and home video screening}.
\bjtitle{Journal of the American Medical Informatics Association}
\bvolume{25}(\bissue{8}),
\bfpage{1000}--\blpage{1007}
(\byear{2018})
\doiurl{10.1093/jamia/ocy039}
\end{barticle}
\endbibitem

\bibitem[\protect\citeauthoryear{Tahmooresi et~al.}{2018}]{tahmooresi2018early}
\begin{botherref}
\oauthor{\bsnm{Tahmooresi}, \binits{M.}},
\oauthor{\bsnm{Afshar}, \binits{A.}},
\oauthor{\bsnm{Rad}, \binits{B.B.}},
\oauthor{\bsnm{Nowshath}, \binits{K.}},
\oauthor{\bsnm{Bamiah}, \binits{M.}}:
Early detection of breast cancer using machine learning techniques
(2018).
\url{https://jtec.utem.edu.my/jtec/article/view/4706}
\end{botherref}
\endbibitem

\bibitem[\protect\citeauthoryear{Gould et~al.}{2021}]{gould2021machine}
\begin{barticle}
\bauthor{\bsnm{Gould}, \binits{M.K.}},
\bauthor{\bsnm{Huang}, \binits{B.Z.}},
\bauthor{\bsnm{Tammemagi}, \binits{M.C.}},
\bauthor{\bsnm{Kinar}, \binits{Y.}},
\bauthor{\bsnm{Shiff}, \binits{R.}}:
\batitle{Machine learning for early lung cancer identification using routine clinical and laboratory data}.
\bjtitle{American Journal of Respiratory and Critical Care Medicine}
\bvolume{204}(\bissue{4}),
\bfpage{445}--\blpage{453}
(\byear{2021})
\doiurl{10.1164/rccm.202007-2791OC}
\end{barticle}
\endbibitem

\end{thebibliography}

\section*{Acknowledgements}

This work was supported by the European Commission’s Horizon Europe Framework Programme under the Research and Innovation Action GA n. 101070546–MUQUABIS, by the European Union’s Horizon 2020 research and innovation programme under FET-OPEN GA n. 828946–PATHOS, by the European Defence Agency under the project Q-LAMPS Contract No B PRJ- RT-989, and by the MUR Progetti di Ricerca di Rilevante Interesse Nazionale (PRIN) Bando 2022 - project n. 20227HSE83 – ThAI-MIA funded by the European Union - Next Generation EU.

G. C. is a member of INFM (INdAM). 

\begin{appendices}

\section{Dataset details}\label{appen}

The 25 handwriting/drawing tasks can be grouped in 3 categories: memory and dictation, graphic, copy. The tasks are described in Table \ref{tab:tasks}.
\begin{table}[h!]
    \caption{List of tasks performed. The tasks are divided in the categories memory and dictation (M), graphic (G), and copy (C).}
    \begin{tabular}{c l c}
    \toprule
\# & Description  & Category \\
\midrule
1 &   Signature drawing & M \\
2 &   Join two points with a horizontal line, continuously for four times & G \\
3 &   Join two points with a vertical line, continuously for four times & G \\
4 &   Retrace a circle (6 cm of diameter) continuously for four times & G \\
5 &   Retrace a circle (3 cm of diameter) continuously for four times & G\\
6 &  Copy the letters ‘l’, ‘m’ and ‘p’  & C\\
7 &    Copy the letters on the adjacent rows & C\\
8 &   Write cursively a sequence of four lowercase letter ‘l’, in a single smooth movement & C\\
9 &   Write cursively a sequence of four lowercase cursive bigram ‘le’, in a single smooth movement & C\\
10 &   Copy the word ‘‘sheet’’ & C\\
11 &  Copy the word ‘‘sheet’’ above a line & C\\
12  &  Copy the word ‘‘mum’’ & C\\
13  &  Copy the word ‘‘mum’’ above a line & C\\
14  &  Memorize the words ‘‘telephone’’, ‘‘dog’’, and ‘‘shop’’ and rewrite them & M\\
15  &  Copy in reverse the word ‘‘bottle’’ & C\\
16  &  Copy in reverse the word ‘‘house’’ & C\\
17  &  Copy six words (regular, non regular, non words) in the appropriate boxes & C\\
18 &   Write the name of the object shown in a picture (a chair) & M\\
19  &  Copy the fields of a postal order & C \\
20 &   Write a simple sentence under dictation & M\\
21 &   Retrace a complex form & G\\
22  &  Copy a telephone number & C\\
23  &  Write a telephone number under dictation & M\\
24  &  Draw a clock, with all hours and put hands at 11:05 (Clock Drawing Test) & G\\
25  &  Copy a paragraph & C\\
\botrule
    \end{tabular}
    
    \label{tab:tasks}
\end{table}
Each task is performed on a different paper sheet and the pile of 25 sheets is put on top of a Wacom’s Bamboo tablet equipped with a pen that can both write in ink on paper and allow the tablet to sample x-y coordinate on paper and on air within a maximum distance of 3cm. The tablet is connected to a PC that processes the raw data to obtain for each task the following features:

\begin{enumerate}
\item Total Time (TT): Total time spent to perform the entire task.
\item Air Time (AT): Time spent to perform in-air movements.
\item Paper Time (PT): Time spent to perform on-paper movements.
\item Mean Speed on-paper (MSP): Average speed of on-paper movements.
\item Mean Speed in-air (MSA): Average speed of in-air movements.
\item Mean Acceleration on-paper (MAP): Average acceleration of on-paper movements. 
\item Mean Acceleration in-air (MAA): Average acceleration of in-air movements.
\item Mean Jerk on-paper (MJP): Average jerk of on-paper movements. Jerk is the variation of acceleration with respect to time.
\item Mean Jerk in-air (MJA): Average jerk of in-air movements.
\item Pressure Mean (PM): Average of the pressure levels exerted by the pen tip.
\item Pressure Var (PV): Variance of the pressure levels exerted by the pen tip.
\item GMRT on-paper (GMRTP): Generalization of the Mean Relative
Tremor (MRT) as defined by Pereira et al. (2015). MRT measures the amount of tremor in drawing spirals and meanders.
\item GMRT in-air (GMRTA): Generalization of the Mean Relative Tremor computed on in air movements.
\item Mean GMRT (GMRT): Average of GMRTP and GMRTA.
\item Pendowns Number (PWN): Counts the total number of pendowns
recorded during the execution of the entire task.
\item Max X Extension (XE): Maximum extension recorded along the X axis. 
\item Max Y Extension (YE): Maximum extension recorded along the Y axis. 
\item Dispersion Index (DI): The Dispersion Index measures how the handwritten trace is ‘‘dispersed’’ on the entire piece of paper. To calculate the index the sheet is ideally divided into TB boxes of 3×3 pixels, then the number CB of boxes containing a fragment of handwriting/drawing is computed. DI is given by the ratio between CB and TB.
\end{enumerate}

\end{appendices}

\end{document}